\documentclass[twocolumn,preprintnumbers,amsmath,amssymb,nobibnotes,nofootinbib]{revtex4}
\usepackage{graphicx}
\usepackage{dcolumn}
\usepackage{color}
\RequirePackage[colorlinks=true
,urlcolor=blue
,anchorcolor=blue
,citecolor=blue
,filecolor=blue
,linkcolor=blue
,menucolor=blue
,pagecolor=blue
,linktocpage=true
,pdfproducer=medialab
]{hyperref}

\begin{document}
%
%

\preprint{FTUAM-12-93} 
\preprint{IFT-UAM/CSIC-12-59}
\preprint{TUM-HEP-844/12}

\title{On the Potential of Leptonic Minimal Flavour Violation}

\author{ R. Alonso}
\email{rodrigo.alonso@uam.es}
\author{M. B. Gavela}
\email{belen.gavela@uam.es}
\affiliation{Departamento de F\'isica Te\'orica, Universidad Aut\'onoma de Madrid and\\
Instituto de F\'{\i}sica Te\'orica IFT-UAM/CSIC, Cantoblanco, 28049 Madrid, Spain}

\author{D. Hern\'andez}
\email{dhernand@ictp.it}
\affiliation{The Abdus Salam International Center for Theoretical Physics,\\
Strada Costiera 11, I-34151 Trieste, Italy}

\author{L. Merlo}
\email{luca.merlo@ph.tum.de}
\affiliation{TUM Institute for Advanced Study, Technische Universit\"at M\"unchen, \\
Lichtenbergstrasse 2a, D-85748 Garching, Germany}

\begin{abstract}
Minimal Flavour Violation can be realised in several ways in the lepton sector due to the possibility of Majorana neutrino mass terms. We derive the scalar potential for the fields whose background values are the Yukawa couplings, for the simplest See-Saw model with just two right-handed neutrinos, and explore its minima. The Majorana character plays a distinctive role: the minimum of the potential allows for large mixing angles -in contrast to the simplest quark case- and predicts 
a maximal Majorana phase. This points in turn  to a strong correlation between neutrino mass hierarchy and mixing pattern.
\end{abstract}
\maketitle

%
%

Fermion families mix under charged-current interactions with a startling different pattern for leptons and quarks: large mixings versus small ones, respectively. The scales of the mass spectrum also differ by many orders of magnitude, with neutrinos in particular exhibiting tiny masses compared to charged leptons and quarks. Furthermore, the charged fermion spectrum is hierarchical, while it remains to be determined whether the neutrino spectrum is hierarchical (normal or inverted) or degenerate. It is plausible that the different mixing patterns may be related to the fact that the neutrino sector may include Majorana mass terms, but this cannot be determined in general.  Nevertheless, we will show that the Minimal Flavour Violation (MFV) ansatz \cite{Chivukula:1987py,Hall:1990ac,D'Ambrosio:2002ex,Cirigliano:2005ck,Davidson:2006bd,Alonso:2011jd} is restrictive enough to allow to explore that question within its framework.

In particular, we will consider the simplest MFV See-Saw model presented in Ref.~\cite{Gavela:2009cd}, that  incorporates only two right-handed (RH) neutrinos (see also Ref.~\cite{Raidal:2004vt}, where scenarios with three RH neutrinos where presented, that reduce in a certain limit to the two light neutrino case). Within a minimalistic approach and in line with the original MFV formulation, each Yukawa coupling will be associated to just one scalar field, usually called flavon, whose vacuum expectation value (vev)  corresponds to the physical masses and mixings. We will show that the study of the associated scalar potential allows to relate the Majorana nature of neutrinos with the large mixings in the lepton sector. This represents a novelty in the field and an advantage with respect to models based on discrete symmetry (see Ref.~\cite{Altarelli:2012ss} and references therein), where usually the mass spectrum and the Majorana phases are not correlated with the mixing angles.

Furthermore, this context allows to explore simultaneously the flavour puzzle for both quarks and leptons,while other studies, such as the so-called ``Anarchy" approach \cite{Hall:1999sn,deGouvea:2003xe,deGouvea:2012ac}, address only the lepton sector. Indeed, the identification of a possible dynamics underlying the flavour structure needs to address both sectors in a comprehensive common framework.

%
%
\section{MFV  See-Saw Model}

The Lagrangian reads 
\begin{equation}
\begin{split}
-\mathcal{L}_{mass}= & \,\overline{\ell}_L\phi Y_E E_R+\overline{\ell}_L\tilde{\phi} \left(Y N+Y' N'\right)+\\
& +\Lambda\overline{N'}N^c+h.c.
\end{split}
 \label{LLN}
\end{equation}
where $\ell_L$ denotes the left-handed (LH) lepton doublet, $E_R$ denotes the RH charged lepton fields (flavour indices are implicit) and $\phi$ denotes the Higgs doublet, with $\tilde \phi=i\tau_2 \phi^*$ and $\langle \phi \rangle \equiv v/\sqrt2$ the electroweak vev, with $v=246$ GeV. The model contains only two heavy neutrinos, $N$ an $N'$,  and in consequence one light neutrino remains massless -an open possibility to this date- and there is only one physical Majorana phase. $\Lambda$ is a Majorana scale, while the Yukawa couplings for charged leptons $Y_E$, and for neutrinos $Y$ and $Y'$, are a matrix and vectors in flavour space, respectively. The Lepton Number (LN) symmetry is violated by the simultaneous presence of $Y$, $Y'$ and $\Lambda$. Indeed, it is sufficient that either $Y$ or $Y'$ vanish to recover LN, as it can be verified with a suitable assignment for the  LN charges of the fields: 
\begin{equation}
\ell_L\rightarrow e^{i\phi}\ell_L\,,\,\,\,
E_R\rightarrow e^{i\phi}E_R\,,\,\,\,
\left(\begin{array}{c}
N\\
N'\\
\end{array}\right)\,\rightarrow\, e^{i\phi\sigma_3}\left(\begin{array}{c}
N\\
N'\\
\end{array}\right)\,,
\label{LNeq}
\end{equation}
where $\sigma_3$ is the third Pauli matrix.  The light neutrino matrix reflects these properties and takes the typical form in type I See-Saw models:  
\begin{equation}
\bar \nu_L\, \frac{v^2}{2\Lambda}\left(Y\,Y'^T+Y'\,Y^T \right) \nu^c_L + h.c.\,,\qquad
\label{eq:lightm}
\end{equation}
where $\Lambda\gg v$ has been assumed.

In this model, flavour changing effects may occur even in the limit of LN conservation, for vanishing  $Y$ or $Y'$ (but not both), and they may be observable for $\Lambda$ values not much larger than the TeV scale~\cite{Gavela:2009cd,Alonso:2010wu,Eboli:2011ia,Dinh:2012bp,AristizabalSierra:2012yy}. 
Without loss of generality, we work in the basis in which the charged lepton Yukawa matrix is diagonal. Denoting by $Y_\nu$  the matrix constructed out of the two vectors $Y$ and $Y'$, 
\begin{equation}
Y_\nu\equiv(Y,Y')\,,
\end{equation}
the Yukawa couplings may be described by~\cite{Gavela:2009cd}
\begin{equation}
Y_E=
 \left(
 \begin{array}{ccc}
  y_e & 0 & 0\\
  0 & y_\mu & 0\\
  0 & 0 & y_\tau\\
  \end{array}
  \right)\,,\quad\!
Y_\nu=U\,f_{m_\nu}\,
\left( \begin{array}{cc}
-iy&iy'\\
y&y'\\
\end{array}\right)\,,
\label{eq:YukawaVevs}
\end{equation}
where $y\equiv\sqrt{ (Y)^\dagger Y}$ and $y'\equiv\sqrt{ (Y')^\dagger Y'}$. Here $U$ denotes the PMNS mixing matrix: using the PDG notation, $U$ is written as the product of three rotations and a matrix containing the Majorana phases, $U=R_{23}(\theta_{23})\,R_{13}(\theta_{13},\delta)\,R_{12}(\theta_{12})\,\Omega$. $\Omega$ can be parametrised as $\Omega=\textrm{diag}\{1, e^{i\alpha}, e^{-i\alpha}\}$ for NH and $\Omega=\textrm{diag}\{ e^{i\alpha}, e^{-i\alpha}, 1\}$ for IH. In the convention that will be used throughout the Letter in which $\Delta_{ij} \equiv m_j^2 -m_i^2>0$, all angles $\theta_{ij}\in[0,\pi/2]$, the Dirac CP phase $\delta\in[0,2\pi)$ and the Majorana phase $\alpha\in[0,\pi]$. The term $f_{m_\nu}$ is a matrix function of neutrino masses: for the normal (NH) and inverted (IH) hierarchies it is defined as 
\begin{equation}
\begin{aligned}
\mbox{NH:}\quad\quad &f_{m_\nu}=\frac{1}{\sqrt{m_{\nu_2}+m_{\nu_3}}}
									\left(\begin{array}{cc}
													0&0\\
													\sqrt{m_{\nu_2}}&0\\
													0&\sqrt{m_{\nu_3}}\\
											\end{array}\right)\,,\\
\left(m_{\nu_2}\right)^2&=\Delta m^2_{sol}\,, \quad
\left(m_{\nu_3}\right)^2=\Delta m^2_{atm}+\Delta m^2_{sol}\,,\\
\mbox{IH:}\quad\quad &f_{m_\nu}=\frac{1}{\sqrt{m_{\nu_1}+m_{\nu_2}}}
									\left(\begin{array}{cc}
													\sqrt{m_{\nu_1}}&0\\
													0&\sqrt{m_{\nu_2}}\\
													0&0\\
											\end{array}\right)\,,\\
\left(m_{\nu_1}\right)^2&=\Delta m^2_{atm}-\Delta m^2_{sol}\,, \quad
\left(m_{\nu_2}\right)^2=\Delta m^2_{atm}\,. 
\label{numasses}
\end{aligned}
\end{equation}
The analysis of the scalar potential is straightforward in the basis in which the heavy singlet neutrino mass matrix is diagonal, in which the Lagrangian in Eq.~(\ref{LLN}) becomes:
\begin{equation}
\begin{split}
-\mathcal{L}_{mass}=&\,\overline{\ell}_L\phi Y_E E_R+\overline{\ell}_L\tilde{\phi}\tilde{Y}_\nu(N_1,N_2)^T+\\
& +\Lambda(\overline{N}_1N^c_1+\overline{N}_2N^c_2)+h.c.\,,
\end{split}
 \label{LCI}
\end{equation}
where the neutrino Yukawa coupling $\tilde Y_\nu$ reads 
\begin{equation}
\tilde{Y}_\nu=\frac{1}{\sqrt{2}}U\,f_{m_\nu}\,
\left(
\begin{array}{cc}
y+y'&-i(y-y')\\
i(y-y')&y+y'\\
\end{array}\right)\,.
\label{eq:YnuRHbasis}
\end{equation}

%
%

\section{The Scalar Potential of MFV}

In the limit of vanishing  Yukawa couplings, $Y_E=\tilde Y_\nu=0$, the Lagrangian in Eq.~(\ref{LCI}) presents an extended global non-Abelian symmetry (plus Abelian global symmetries which are not made explicit below),
\begin{equation}
\begin{split}
\mathcal{G}_{fl}
\sim SU(3)_{\ell_L}\times SU(3)_{E_R}\times O(2)_N\,,
\end{split}
\end{equation}
with $O(2)_N$ corresponding to orthogonal transformations between $N_1$ and $N_2$ (in the flavour basis, the Majorana mass term in Eq.~(\ref{LLN}) exhibits a $U(1)_N\times Z_2$ invariance in the $(N,N')$ sector, isomorphic to $O(2)_N$).
Notice that when $\mathcal{G}_{fl}$ is exact, LN is also a symmetry of the Lagrangian, according to Eqs.~(\ref{LLN}) and (\ref{LNeq}). 
 For non-zero  Yukawa couplings the flavour symmetry is formally recovered promoting $Y_E$ and $\tilde Y_\nu$ to dimensionless auxiliary fields transforming under $\mathcal{G}_{fl}$ as follows:
\begin{equation}
Y_E\,\sim\, (\,3\,,\,\bar 3\,,\,1)\,,\qquad \tilde Y_\nu\,\sim\, (\,3\,,\,1\,,\,2)\,, 
\label{spurions}
\end{equation}
for the charged lepton and the neutrino sectors, respectively. Given the symmetry $\mathcal{G}_{fl}$ and being $Y_E$ and $Y_\nu$ the only sources of flavour violation, possible diagonal Majorana mass terms are forbidden.

MFV suggests a dynamical origin for the Yukawa couplings of the theory. For instance, they may result from scalar fields, usually called flavons, taking a vev. In the simplest realisation of MFV, each Yukawa coupling is associated to a single scalar field, singlet under  the SM gauge group and transforming under $\mathcal{G}_{fl}$ as the spurions in Eq.~(\ref{spurions}):
\begin{equation}
\mathcal{Y}_E\sim (\,3\,,\,\bar 3\,,\,1)\,,\qquad \mathcal{Y}_\nu \sim (\,3\,,\,1\,,\,2)\,.
\end{equation}
The charged-lepton flavon fields $\mathcal{Y}_E$ belong then to the bi-fundamental representation of $SU(3)_{\ell_L}\times SU(3)_{E_R}\subset {G}_{fl}$, while the neutrino flavon $\mathcal{Y}_\nu$ belongs to the fundamental representation of $SU(3)_{\ell_L}$.

 The relation between the Yukawa couplings and the flavon vevs may be linear,  
\begin{equation}
\frac{\left\langle \mathcal{Y}_E\right\rangle}{\Lambda_{fl}}\equiv Y_{E}\,,\quad
\frac{\left\langle\mathcal{Y}_\nu\right\rangle}{\Lambda_{fl}}\,\equiv \tilde Y_\nu\,,
\label{eq:FlavonsVevs}
\end{equation}
where $\Lambda_{fl}$ represents the scale of the flavour dynamics.  Alternatively, those relations could be non-linear, for instance flavon vevs and Yukawas could be inversely proportional to each other. The analysis of the mixing invariants below will apply to both cases, as it only depends on the field transformation properties. For concreteness, we will stick in this letter to the identification in Eq.~(\ref{eq:FlavonsVevs}). With this assignment, the Yukawa terms in Eq.~(\ref{LCI}) have dimension 5; they are the leading non-trivial flavour invariant operators.

The scalar potential for the  $\mathcal{Y}_E$ and $\mathcal{Y}_\nu$ fields must 
 be invariant under the SM gauge symmetry and the flavour symmetry $\mathcal{G}_{fl}$. 
At the renormalisable level,  the possible linearly-independent invariant terms reduce to:
\begin{equation}
\begin{gathered}
\mbox{Tr}\left(\mathcal{Y}_E\mathcal{Y}_E^\dagger\right)\,, \quad
\mbox{Tr}\left(\mathcal{Y}_\nu \mathcal{Y}_\nu^\dagger\right)\,,\quad
\det\left(\mathcal{Y}_E\right)\,,\\
\mbox{Tr}\left(\mathcal{Y}_E\mathcal{Y}_E^\dagger\right)^2\,,\quad
\mbox{Tr}\left(\mathcal{Y}_E\mathcal{Y}_E^\dagger\mathcal{Y}_\nu\mathcal{Y}_\nu^\dagger\right)\,,\\
\mbox{Tr}\left(\mathcal{Y}_\nu\mathcal{Y}_\nu^\dagger\right)^2\,,\quad
\mbox{Tr}\left(\mathcal{Y}_\nu\sigma_2 \mathcal{Y}_\nu^\dagger\right)^2\,,
\end{gathered}
\end{equation}
leading to the following scalar potential $V$  for the flavon fields:
\begin{align}
V=&-\mu^2\cdot\mathbf{X}^2+\left(\mathbf{X}^2\right)^\dagger\,\lambda\,\mathbf{X}^2+
\left(\mu_D\det\left(\mathcal{Y}_E\right)+h.c.\right)+\nonumber\\
& +\lambda_E\,\mbox{Tr}\left(\mathcal{Y}_E\mathcal{Y}_E^\dagger\right)^2+
g\,\mbox{Tr}\left(\mathcal{Y}_E\mathcal{Y}_E^\dagger\mathcal{Y}_\nu\mathcal{Y}_\nu^\dagger\right)+
\label{potential}\\
&+h\, \mbox{Tr}\left(\mathcal{Y}_\nu\mathcal{Y}_\nu^\dagger\right)^2+
h'\, \mbox{Tr}\left(\mathcal{Y}_\nu\sigma_2 \mathcal{Y}_\nu^\dagger\right)^2\,.\nonumber
\end{align}
In this equation $\mathbf{X}^2$ is a two-component vector defined by
\begin{equation}
 \mathbf{X}^2\equiv\left(
                     \mbox{Tr}\left(\mathcal{Y}_E\mathcal{Y}_E^\dagger\right)\,,
					\mbox{Tr}\left( \mathcal{Y}_\nu^\dagger \mathcal{Y}_\nu\right)
					\right)^T\,,
\nonumber
\end{equation}
$\mu^2$ is a real two-component vector, $\lambda$ is a $2\times 2$ Hermitian matrix and all other coefficients  are real parameters, except for $\mu_D$ which may be complex. The full scalar potential includes in addition  Higgs-$\mathcal{Y}_E$ and Higgs-$\mathcal{Y}_\nu$ cross-terms, but they do not affect the mixing pattern and will thus be obviated in what follows.

The question now is whether the minimum of the potential can accommodate the physical values of masses and mixings and whether its coefficients need to be fine-tuned. For charged leptons, the choice of  $\mathcal{Y}_E$ in the bifundamental mirrors that  for the simplest quark case~\cite{Anselm:1996jm,Berezhiani:2001mh,Feldmann:2009dc,Alonso:2011yg}. It is worth noting that the determinant $\det\left(\mathcal{Y}_E\right)$, whose impact is to push towards degenerate charged lepton masses, is 
absent when the $\mathcal{G}_{fl}$ symmetry is enlarged to include the invariance under global phases, 
\begin{equation}
\mathcal{G}_{fl}= U(3)_{\ell_L} \times U(3)_{E_R} \times O(2)_N\,,
\label{NewFlavourSym}
\end{equation}
present in the limit of vanishing Yukawa couplings. Furthermore, with this enlarged symmetry, the first non-renormalisable contributions to the scalar potential arise only at mass dimension 6, and would be suppressed by $\Lambda^2_{fl}$.

For the $\mathcal{Y}_\nu$ fields, an interesting issue is whether the potential minimum allows for the hierarchy $y\gg y'$ or $y'\gg y$, or in other words the approximate LN invariant case. The answer is positive. It is useful at this point to consider $\mathcal{Y}_\nu$ as a composition of two fields,
\begin{equation}
\mathcal{Y}_\nu\equiv \left(\dfrac{i}{\sqrt2}\left(\mathcal{Y}-\mathcal{Y}'\right),\,\dfrac{1}{\sqrt2}\left(\mathcal{Y}+\mathcal{Y}'\right)\right)
\end{equation}
where $\mathcal{Y}$ and $\mathcal{Y}'$ transform as triplets of $SU(3)_{\ell_L}$ and under $U(1)_N\times Z_2$. Then,  the minimum of the scalar potential in Eq.~(\ref{potential}) implies that $\mathcal{Y}^\dag\mathcal{Y}$ and $\mathcal{Y}'^\dag\mathcal{Y}'$ lie on a circle, while  the aforementioned hierarchy is driven by the following terms:
\begin{equation}
\begin{split}
V\supset-\left(h+h'-\left(h-h'\right)\cos^2\Delta \right)\,\left( \mathcal{Y}^\dagger \mathcal{Y}\right)\left(\mathcal{Y}'^\dagger \mathcal{Y}'\right)\,,
\end{split}
\end{equation}
where $\cos\Delta$ encodes the relative alignment in the flavour space of $\mathcal{Y}$ and $\mathcal{Y}'$ fields,
\begin{equation}
\begin{aligned}
\cos \Delta = \dfrac{m_{\nu_3} - m_{\nu_2}}{m_{\nu_3} + m_{\nu_2}}  \quad & \text{for the NH}\\
\cos \Delta = \dfrac{m_{\nu_2} - m_{\nu_1}}{m_{\nu_2} + m_{\nu_1}} \quad & \text{for the IH}\,.
\end{aligned}
\end{equation}
For negative values of $\left(h+h'-\left(h-h'\right)\cos^2\Delta \right)$, the minimum 
corresponds to either $y$ or $y'$ vanishing and thus to LN conservation; while $y=y'$ is favoured for positive values. 

In the spirit of an effective field theory approach, the analysis presented here consistently considers all possible operators up to mass dimension 5, invariant under the SM gauge symmetry and the flavour symmetry in Eq.~(\ref{NewFlavourSym}). The first relevant contributions to the Yukawa lagrangian appear at dimension 5, while the scalar potential is non-trivial already at the renormalizable level, \mbox{dim $\le4$}. The next to leading contributions appear in both sectors at mass dimension 6 and have a negligible impact on the determination of the mixing angles and the Majorana phase, that will be discussed in the following section. These contributions will thus not be further considered.

%
%

\subsection{Mixing angles and Majorana Phase}

Consider  now the fermion masses fixed at their physical values and focus on the mixing pattern allowed at the minimum of the potential. Since  mixing  arises from the misalignment in flavour space of the charged lepton and the neutrino flavons, the only relevant invariant at the renormalisable level  is
\begin{equation}
\mathcal{O}_{\textrm{mix}}\equiv\mbox{Tr}\left(\mathcal{Y}_E\mathcal{Y}_E^\dagger\mathcal{Y}_\nu\mathcal{Y}_\nu^\dagger\right)\,.
\label{Mixing-trace}
\end{equation}
Substituting the expressions for the flavon vevs in Eq.~(\ref{eq:FlavonsVevs}), it follows that:
\begin{widetext}
\begin{equation}
\mathcal{O}_{\textrm{mix}}=\frac{2\Lambda_{fl}^4}{v^2\sum m_{\nu_i}}\left(( y^2+ y'^2)\sum_{l,i} |U^{li}|^2m_l^2 m_{\nu_i}+ 
(y^2- y'^2)\left(i\,e^{2i\alpha}\,\sum_{l,i<j}(U^{li})^*U^{lj}m_l^2\sqrt{m_{\nu_i} m_{\nu_j}} +c.c.\right)\right)\,.
\label{genericmixing}
\end{equation}
\end{widetext}
This expression can be compared with the equivalent one for quarks in the bifundamental of the flavour group~\cite{Alonso:2011yg}, given by
\begin{equation}
\mbox{Tr}\left(\Sigma_u\Sigma_u^\dagger\Sigma_d\Sigma_d^\dagger\right)=\frac{4\Lambda_{fl}^4}{v^4}  \sum_{i,j} |U_{CKM}^{ij}|^2m_{u_i}^2 m_{d_j}^2\,,
\label{quarkmixing}
\end{equation}
in an obvious notation. The first term in Eq.~(\ref{genericmixing}) for leptons corresponds to that for quarks in Eq.~(\ref{quarkmixing}): the only difference is the linear -instead of quadratic- dependence on neutrino masses, as befits the See-Saw realisation. The second term in Eq.~(\ref{genericmixing}) has a strong impact on the localisation of the minimum of the potential and is responsible for the different results in the quark and lepton sectors: it contains the Majorana phase $\alpha$ and therefore connects the Majorana nature of neutrinos to their mixing.

%
%
\subsection*{Two lepton families}

In the illustrative two generation case, in which the flavour symmetry is $\mathcal{G}_{fl}= SU(2)_\ell \times SU(2)_{E_R} \times O(2)$ (plus Abelian symmetries), the mixing term in the potential Eq.~(\ref{potential}) and Eq.~(\ref{genericmixing}) can be written as follows:
\begin{equation}
\begin{split}
g\mathcal{O}_{\textrm{mix}}\propto
&g\Big\{(m_e^2+m_\mu^2)(y^2+y'^2)(m_{\nu_2}+m_{\nu_1})+\\ 
&+(m_\mu^2-m_e^2)\Big[ (m_{\nu_2}-m_{\nu_1})(y^2+y'^2)\cos2\theta+\\ 
&+(y^2-y'^2)2\sqrt{m_{\nu_2}m_{\nu_1}}\sin2\alpha\sin2\theta\Big] \Big\}\,,
\end{split}
\label{two-family-pot}
\end{equation}
where $\theta$ is the mixing angle and $\alpha$ the Majorana phase.
 This formula shows explicitly the relations expected on physical grounds, between the mass spectrum and non-trivial mixing: i) the dependence on the mixing angle disappears in the limit of degenerate charged lepton masses; ii) it also vanishes for degenerate neutrino masses if and only if $\sin2\alpha=0$; iii) on the contrary, for  $\sin2\alpha\ne 0$ the dependence on the mixing angle remains, as it is physical even for degenerate neutrino masses;  iv) the $\alpha$ dependence vanishes when one of the two neutrino masses vanishes or in the absence of mixing, as $\alpha$ becomes then unphysical.

The minimisation with respect to the Majorana phase and the mixing angle leads to the constraints:
\begin{gather}
(y^2-y'^2)\sqrt{m_{\nu_2}m_{\nu_1}}\sin2\theta\cos2\alpha=0\,,
\label{cos2alpha}\\
\mbox{tg}2\theta=\sin2\alpha\frac{y^2-y'^2}{y^2+y'^2}
\frac{2\sqrt{m_{\nu_2}m_{\nu_1}}}{m_{\nu_2}-m_{\nu_1}}\,.
\label{tan2theta}
\end{gather}
The convention used implies that  $\sin\theta_{ij}>0$ and $\cos\theta_{ij}>0$. The first condition predicts then that the {\it Majorana phase is maximal, $\alpha=\{\pi/4,3\pi/4\}$, for non-trivial mixing angle}. The relative Majorana phase between the two neutrinos is therefore $2\alpha=\pm\pi/2$ which implies no CP violation due to Majorana phases. On the other hand, Eq.~(22) establishes  a link between the mixing strength and the type of spectrum, which indicates {\it a maximal angle for degenerate neutrino masses, and  a small angle for strong mass hierarchy.} The sign of $\cos2\theta$ is selected by the absolute minimum of the potential, which depends on the sign of $g$, see Eq.~(\ref{two-family-pot}): $\cos2\theta$ negative (positive) for $g>0$ ($g<0$). 
 The sign of the last term in Eq.~(\ref{two-family-pot}) in turn determines the sign of $\sin2\alpha$ for given $y,y'$.

Using Eqs.~(\ref{cos2alpha}) and (\ref{tan2theta}), at the minimum of the potential, the mixing invariant in Eq.~(\ref{Mixing-trace}) acquires a simple interpretation in terms of the eigenvalues of $\mathcal{Y}_E\mathcal{Y}_E^\dagger$ and  $\mathcal{Y}_\nu\mathcal{Y}_\nu^\dagger$, which are respectively proportional to the charged lepton masses  $m_{l_i}$, and to the combinations $m_\pm$ as follows:
\begin{equation}
\begin{cases}
\mathcal{O}_{\textrm{mix}}\Big|_{min}\propto m_e^2\, m_+ +m_\mu^2\,m_-\,,&\quad g>0\,,\\[2mm]
\mathcal{O}_{\textrm{mix}}\Big|_{min}\propto m_e^2\, m_- +m_\mu^2\,m_+\,,&\quad g<0\,,
\end{cases}
\label{trace-minimum}
\end{equation}
where
\begin{align}
&m_\pm \equiv a_\nu\pm\sqrt{a_\nu^2-c_\nu^2}\,,\label{eq:Defmpm}   \\
&a_\nu=(m_{\nu_2}+m_{\nu_1})(y^2+y'^2)\,,\,\, c_\nu=4\sqrt{m_{\nu_2}m_{\nu_1}}yy'\,.\nonumber
\end{align}
Depending on the sign of $g$, the two set of eigenvalues, $m_{l_i}$ and $m_\pm$, are thus combined in a sum of their products in increasing or decreasing mass order (more generally, these results are a consequence of the Von Neumann's trace inequality).

As an illustrative exercise within the two-family scenario, let us consider the ``solar" $(1,2)$ sector: can the observed value of  $\theta_{12}$ be accounted for in this framework? The answer is affirmative: $\cos2\theta^{exp}_{12}>0$ and $\sin2\theta^{exp}_{12}>0$ and the scalar potential can accommodate these facts at its minimum for $g<0$; if $y>y'$ ($y<y'$) then $\tan 2\theta_{12}>0$ for $\alpha=\pi/4$ ($\alpha=3\pi/4$), see  Eq.~(\ref{tan2theta}).

This simple line of reasoning has the advantage that it extends straightforwardly to the three lepton generation case, 
which in this model adds a massless neutrino and thus a vanishing third $\mathcal{Y}_\nu\mathcal{Y}_\nu^\dagger$ eigenvalue.

%
%
\subsection*{Three lepton families}
%

The model under discussion deals with two light massive neutrinos, a massless one and only one Majorana phase. For the angle and Majorana phase spanned by the two light massive eigenstates, Eqs.~(\ref{cos2alpha}) and (\ref{tan2theta}) apply. Nevertheless, the other two mixing angles are forced to vanish by construction, and this minimal version of the model is not viable. It is interesting though to further analyse more the source of the problem, as a guideline for a realistic three family scenario.

The two-generation expression in Eq.~(\ref{trace-minimum}) for the mixing invariant at the minimum of the potential is now replaced by: 
\begin{equation}
\begin{cases}
\mathcal{O}_{\textrm{mix}}\Big|_{min}\propto m_e^2\, m_+ +m_\mu^2\,m_-&\quad g>0\\[2mm]
\mathcal{O}_{\textrm{mix}}\Big|_{min}\propto m_\mu^2\, m_- +m_\tau^2\,m_+&\quad g<0
\end{cases}
\label{trace-minimum_3fam}
\end{equation}
where $m_{\pm}$ are defined as in Eq.~(\ref{eq:Defmpm}) except that the mass eigenvalues entering the definition of $a_\nu$ and $c_\nu$  are hierarchy dependent: $m_{\nu_1,\nu_2}$ hold for IH (which requires $g>0$) while they are substituted  by $m_{\nu_2,\nu_3}$ for NH ($g<0$). It is easily verified that for NH the resulting atmospheric angle would be small, in contradiction with data. For IH, although the resulting solar angle would be large, the minimum of the potential forces it to lie in the second octant, again a possibility excluded by data. This shows the strong constraining power of the minimisation procedure.  
 
In consequence, for the simple See-Saw model and flavon assignment considered here, the mixing pattern at the minimum of the potential is not compatible with that observed in nature. One source of the problem is the value of the one mixing angle obtained, the other problem is the absence of the other two angles.

For the value of the mixing angle obtained a word of caution is pertinent, though. Although the number and kind of invariants in the potential is only dictated by the gauge and flavour symmetries,  the spectrum of masses is more model-dependent. For instance, in models in which the light lepton masses are inversely proportional to the flavon vevs (instead of linearly proportional as assumed above), the analysis of the three-family case may allow a realistic identification of the non-vanishing angle with the solar one and point to an IH pattern, for $g<0$. For instance, a context in which 
 $\langle\mathcal{Y}_E\rangle\propto \text{diag}(1/y_e,\,1/y_\mu,\,1/y_\tau)$ would be alike to that in See-Saw scenarios which include heavy fermions.  This feature is naturally embedded in models where the flavour symmetry is gauged \cite{Grinstein:2010ve,Feldmann:2010yp,Guadagnoli:2011id}: in order to guarantee anomaly cancellation, heavy fermions enrich the spectrum; when flavons (or spurions in simpler scenarios) enter only the heavy sector and the portal between the heavy and the SM sectors is flavour blind, then the SM Yukawas result inversely proportional to the flavon vevs.

%
%
\section{Conclusions}

An outstanding question is whether the strong mixing pattern found in the leptonic sector - in contrast to the small mixing and hierarchical spectrum of the quark sector - is related to the possible Majorana character of the neutrino fields. While this cannot be answered in general, we have shown that the hypothesis of a dynamical realisation of MFV  may be restrictive enough to answer it within its framework.

We have explored the possibility of a dynamical origin of the Yukawa couplings of leptons in the context of MFV and Majorana neutrinos.
The simplest realisation  is to identify the Yukawa couplings with the vevs of some dynamical scalar fields, the flavons. The implementation of the Majorana character needs to refer to an explicit model of Majorana neutrino masses. 
 The potential can then be determined in general, under the only requirement of SM gauge invariance and invariance under the underlying flavour symmetry.   Much as it was the case for the analogous analysis for quarks~\cite{Anselm:1996jm,Berezhiani:2001mh, Alonso:2011yg}, the latter turns out to be strongly restrictive. This Letter contains  the first analysis of this type when Majorana neutrinos are present. 

We  concentrated here in one of the simplest possible MFV models of Majorana neutrino masses: a See-Saw model  with only two extra heavy singlet neutrinos with approximate $U(1)$ lepton symmetry \cite{Gavela:2009cd}. We determined the corresponding scalar potential and explored its minima. 

While for quarks the minimal assumption of associating each Yukawa coupling to just one flavon led to the conclusion of no mixing for a renormalisable potential~\cite{Anselm:1996jm,Berezhiani:2001mh,Feldmann:2009dc,Alonso:2011yg}, the presence of the Majorana character introduces radically new ingredients. Generically, the flavour group is enlarged and different invariants are allowed in the potential. The toy model with only two families demonstrates that non-trivial Majorana phases and mixing angles may be selected by the potential minima and indicates a novel connection with the pattern of neutrino masses: i) large mixing angles are possible; ii) there is a strong correlation between mixing strength and mass spectrum; iii) the relative Majorana phase among the two massive neutrinos is predicted to be maximal, $2\alpha= \pi/2$, for non-trivial mixing angle; moreover, although the Majorana phase is maximal, it does not lead to CP violation, as it exists a basis in which all  terms in the Lagrangian are real. 
These novel results are intimately related with the well-known fact that, in the presence of non-trivial Majorana phases, there may be physical mixing even for degenerate neutrino masses.  In consequence, the results are expected to hold as well for general fermionic See-Saw scenarios.
  
For the specific model considered, the Yukawa couplings are linearly proportional to the vev of the flavons, and maximal mixing turns out to be correlated with degenerate neutrino masses for the two massive neutrino eigenstates, while small mixing is correlated with strong mass hierarchy. As a result, the observed ``solar" mixing pattern is well accommodated at the minimum of the potential, but only when restricting the analysis of the model to a two-family approximation.  

For three lepton generations and under the same model assumptions, the minimum of the potential cannot correspond to the measured values of masses and mixings. This is probably linked to the necessary masslessness of the third left-handed neutrino in the minimal model, which imposes a fixed hierarchy with respect to the massive modes. More freedom is expected in models with three massive neutrinos.

The possibility of accommodating large mixing angles and Majorana phases is reminiscent of the typical anarchy pattern~\cite{Hall:1999sn,deGouvea:2003xe,deGouvea:2012ac}, in which the probability peaks at maximum values of the phases. Nevertheless, the present approach goes beyond anarchy not only in that it has dynamical content, but in that it sheds light on the difference in mixing strength in the quark and lepton sectors, while anarchy cannot deal with the quark sector. Furthermore, the model discussed here has an advantage with respect to models based on discrete symmetry, because usually in the latter the mass spectrum and the Majorana phases are not correlated to the mixing angles. 
The strong correlation found between the neutrino mass spectrum, the mixing strength and the Majorana phase is also expected to hold for general fermionic See-Saw models, being linked only to the kind of invariants entering the scalar potential. We will investigate this possibility in a separate paper.

%
%
\begin{acknowledgments}
R. Alonso and M.B. Gavela acknowledge partial support from the  European Union FP7  ITN INVISIBLES (Marie Curie Actions, PITN- GA-2011- 289442). R. Alonso and M.B. Gavela also acknowledge CiCYT support through the project FPA2009-09017, CAM partial support through the project HEPHACOS P-ESP-00346, and partial support from the  European Union FP7 ITN UNILHC (Marie Curie Actions, PITN-GA- 2009-237920). R. Alonso acknowledges MICINN support through the grant BES-2010-037869. L. Merlo acknowledges the Te\-ch\-ni\-sche Universit\"at M\"unchen -- Institute for Advanced Study, funded by the German Excellence Initiative. We thank the Galileo Galilei Institute for Theoretical Physics for the hospitality and the IFNF for partial support during the completion of this work.
\end{acknowledgments}

%

\providecommand{\href}[2]{#2}\begingroup\raggedright\endgroup

\end{document}